# Rapid Reconstruction of 3-D Membrane Pore Structure Using a Single 2-D Micrograph

Hooman Chamani[1], Arash Rabbani[2], Kaitlyn P. Russell[3], Andrew L. Zydney[3], Enrique D. Gomez[3,4], Jason Hattrick-Simpers[5]*, Jay R. Werber[1]*

[1]Department of Chemical Engineering & Applied Chemistry, University of Toronto, Ontario M5S 3E5, Canada.
[2]School of Computing, University of Leeds, Leeds LS2 9JT, UK.
[3]Department of Chemical Engineering, The Pennsylvania State University, Pennsylvania 16802, USA.
[4]Department of Materials Science & Engineering, The Pennsylvania State University, Pennsylvania 16802, USA.
[5]Department of Materials Science & Engineering, University of Toronto, Ontario M5S 3E4, Canada.

## Abstract

Conventional 2-D scanning electron microscopy (SEM) is commonly used to rapidly and qualitatively evaluate membrane pore structure. Quantitative 2-D analyses of pore sizes can be extracted from SEM, but without information about 3-D spatial arrangement and connectivity, which are crucial to the understanding of membrane pore structure. Meanwhile, experimental 3-D reconstruction via tomography is complex, expensive, and not easily accessible. Here, we employ data-science tools to demonstrate a proof-of-principle reconstruction of the 3-D structure of a membrane using a single 2-D image pulled from a 3-D tomographic data set. The reconstructed and experimental 3-D structures were then directly compared, with important properties such as mean pore radius, mean throat radius, coordination number and tortuosity differing by less than 15%. The developed algorithm will dramatically improve the ability of the membrane community to characterize membranes, accelerating the design and synthesis of membranes with desired structural and transport properties.

*Corresponding authors:
J. Hattrick-Simpers (E-mail: jason.hattrick.simpers@utoronto.ca, Tel.: +1-416-978-3012)
J. Werber (E-mail: jay.werber@utoronto.ca, Tel.: +1-416-978-4906)

Introduction

Porous membranes are vital in advanced fields like catalysis, biopharmaceutical separations, batteries, biosensing, microfluidics, air filtration, and water treatment (1). Porous membranes are thin layers (typically < 100 µm) in which the pores are engineered such that the membrane has high flux and selectivity, while tolerating the required range of operating pressures and temperatures. Characterization of porous membranes in terms of key structural parameters, such as pore size distribution (PSD) and throat size distribution (TSD), is imperative to fundamentally understand structure/transport/performance relationships (2). For instance, knowing PSD and TSD permits the evaluation of how particle retention takes place in membranes used for separation processes, as the throat and small pores are the locations where particles are captured.

Different physical techniques have been presented for estimation of pore size and PSD including gas-liquid porometry, liquid-liquid porometry, liquid-vapor equilibrium method, gas-liquid equilibrium method (permporometry), liquid-solid equilibrium method (thermoporometry), and mercury porosimetry. Among these methods, the gas-liquid and liquid-liquid porometry techniques have been most frequently employed (3). In these approaches, the membrane is wetted with a liquid, which is then displaced using gas for membranes with large pores (> 500-nm diameter) or an immiscible liquid with low surface tension for membranes with smaller pores (as small as 2-nm diameter). The obtained flow rate vs. applied pressure curves are then used for the estimation of PSD. Although these methods are quick, reproducible, and affordable, they have drawbacks, including error associated with uncertainties in the contact angle (3–6), assumption of circular cylinders (3), and the possibility of membrane compression during testing. More crucially, these methods require modeling membranes as an array of discrete, parallel pores, which does not suitably capture the anisotropic, complex, and highly connected nature of real membranes.

Scanning electron microscopy (SEM) and transmission electron microscopy (TEM) are methods used frequently to qualitatively characterize the pore size, pore shape, morphology, and structure of membranes (2, 7–10). SEM is used routinely, both for analysis of membrane surface pores and for analysis of membrane cross-sections to assess the overall pore structure. However, quantitative analysis of electron micrographs is uncommon and typically limited to 2-D images, which can yield properties such as surface pore size distribution, surface porosity, overall porosity, total membrane thickness, and selective layer thickness, but cannot directly yield information about the actual 3-D pore size distributions, tortuosity, spatial pore arrangement, and connectivity of pores. Quantitative analysis of surface pores from top-view 2-D images requires the assumption that the minimum pore size occurs directly at the upper surface, which is insufficient for complete modeling and not true in many cases (2). The limitations of analyzing membranes using 2-D images typically necessitates reconstructing the 3-D structure of the membrane for comprehensive analysis.

Serial sectioning is a proven experimental technique for obtaining 3-D information from methods that conventionally produce 2-D data. The technique is composed of two primary steps, (i) iteratively sectioning a material at desired depth increments and (ii) collecting data from each section. Sectioning can be done by several different methods, including polishing, cutting, laser ablation, grinding, and ion sputtering (11). Sectioning and imaging are repeated until the required material volume is interrogated. Following the completion of the experimental data collection, 3-D structures are rebuilt and examined through image-processing and visualization programs. Serial sectioning tomography experiments may now be scaled down to analyze features at the nanometer scale because of the development of focused ion beam (FIB) and FIB-SEM instruments (12). Other techniques for 3-D pore visualization, particularly X-ray computed tomography, have been used for 3-D membrane reconstruction (13, 14); however, given that X-ray computed tomography is limited to resolutions typically > 1 μm, FIB-SEM is preferred over other techniques for membrane characterization owing to its resolution approaching 1 nm (15).

In membrane science, FIB-SEM has recently enabled comprehensive analysis of structure/performance relationships. The technique has been employed to examine microfiltration (16, 17), ultrafiltration (18), reverse osmosis (19), and viral filtration membranes (20). In these studies, 3-D reconstructions of membrane structure were typically obtained with a resolution of 3–5 nm, allowing for calculation of porosity, average pore size, distribution of pores and throat sizes, and percentage of dead-end pores. Also, it made it possible to see how pores in various membrane regions were connected to one another. Critically, fluid flow in the obtained 3-D pore structures can be modeled exactly using computational fluid dynamics (CFD), enabling insight into performance metrics such as the deposition of particles (20), water permeability (20), and air flow rate (21).

Despite its clear benefits, generating 3-D tomographic images from FIB-SEM remains challenging for soft, poorly conducting materials such as polymeric membranes, which are sensitive to ion and electron beam damage. Typical challenges include charging, cross sectioning artifacts, and shadowing impacts, which take place when surrounding materials shadow a pore (22). In addition, FIB-SEM is expensive, requires months of optimization, and is not easily accessible, leading to just nine studies that we are aware of on membrane 3-D reconstructions using FIB-SEM (16–20, 23–26). The time-consuming nature of FIB-SEM also limits sample volumes, which are typically on the order of 1–10 $\mu m^3$ and can only focus on a small part of the membrane (e.g., the region near the membrane inlet or outlet). Given the rapid innovation in membrane science occurring to address sustainability and technological challenges, there is a clear need for an accessible and rapid method for generating representative 3-D pore structures.

In this study, we use data science to create statistically representative 3-D membrane structures using single 2-D images. To develop the algorithm, we used a recently published tomographic FIB-SEM data set (20) from an ultrafiltration membrane used for viral filtration, which enabled us to (i) pull individual 2-D cross-sectional SEM images as inputs to the algorithm and (ii) assess the validity of the algorithm by comparing the experimental and reconstructed 3-D structures. During the optimization of the digitally-created structure, the algorithm only considers a single 2-D input image, with statistical features of the input image compared with a 2-D slice of the digitally-created structure. The structural and performance properties of the obtained 3-D structure were then compared with those calculated from the original reconstruction. With further optimization, the method could revolutionize the structural characterization of membranes, enabling rapid and inexpensive generation of rich 3-D-based data sets to accelerate membrane development.

## Results

### Image segmentation

To develop the algorithm, we used a published FIB-SEM data set (Data S1) of the Viresolve® Pro from MilliporeSigma (20), a commercial membrane that is used in bioprocessing to remove putative viruses as small as 20 nm in diameter from therapeutic proteins; this is typically classified as an ultrafiltration membrane. In this data set, there were 400 cross-sectional SEM images (slices) taken sequentially at 3-nm increments. We have oriented the data set such that the electron micrographs were imaged sequentially in the y-direction, with the z-axis describing depth through the membrane and the x-y plane at z=0 being the top surface (Fig. S5).

The algorithm starts by selecting a random slice and performing image segmentation, in which a label is assigned to each pixel of the image such that pixels with the same label have similar visual characteristics (27). The easiest segmentation approach is thresholding (28). When segmenting an image using the thresholding technique, any pixels whose intensity levels exceed the threshold are fixed to a foreground value; the other pixels will have a background value. Using this method, a grayscale image can be converted into a binary image (29). In the previous study (20) in which FIB-SEM was used to reconstruct the 3-D structure of the membrane, a conventional thresholding method was employed in which a global threshold is applied to all pixels. As can be seen in Fig. S1B, the conventional thresholding method was not fully successful in identifying pore-solid boundaries. Here, we utilized adaptive thresholding, where the threshold value dynamically changes at different parts of the image. The adaptive thresholding algorithm in MATLAB (30) was used, in which the local mean intensity around each pixel is used to determine the threshold value (Fig. S1C).

### Feature extraction from 2-D images

In the computer-vision field, a "feature" is defined as a measurable piece of data that is extracted from the image. Working with features of the image, instead of the image, decreases information redundancy and reduces computational intensity during model training. A challenging topic in computer vision is to extract information-rich features that capture the intrinsic content of the images (31). For instance, in this study, the segmented image selected from the FIB-SEM data has 899×1024 pixels, i.e., 920,576 elements. We reduced the number of elements from 920,576 to 100 by using a histogram of the distance map as the feature. In a previous study, Rabbani et al. (32) calculated the Lattice Boltzmann Method-based absolute permeability of throats and compared the calculated permeabilities with various features extracted from images of porous structures (in this case porous rocks), including cross-sectional area, axes ratio, equivalent radius, solidity (area of the throat convex hull divided by the throat area), and mean distance map. It was observed that the averaged distance map of the throat (throat refers to constrictions between pores, see Fig. S2) was highly correlated with water flow (32), suggesting the usefulness of distance maps for analysis of porous structures. Distance maps have also been employed in other fields including prediction of thermal conductivity of woven ceramic composites (33), binarization of degraded document images (34), fingerprint matching (35), and tumor characterization (36).

Here, we employed a novel distance map to capture the statistical distribution of the pore and solid through a membrane. After image segmentation, 0s were assigned to void (pore) spaces and 1s were assigned to solid spaces. The Euclidean distance between each pixel and the nearest non–zero pixel was calculated (Fig. S3). Instead of the averaged distance map used by Rabbani et al. (32), we used a normalized histogram of the distance map. Fig. 1 (top of the figure) illustrates the steps used to develop a normalized histogram of the distance map for a synthetic binary image. Each histogram was normalized to the total number of pixels in the image. To generate the normalized histogram of the distance map for the void space, similar calculations as those used for the solid phase were repeated (bottom of Fig. 1). Afterward, we combined the normalized histograms to yield a single combined distance function (CDF) (Fig. 1).

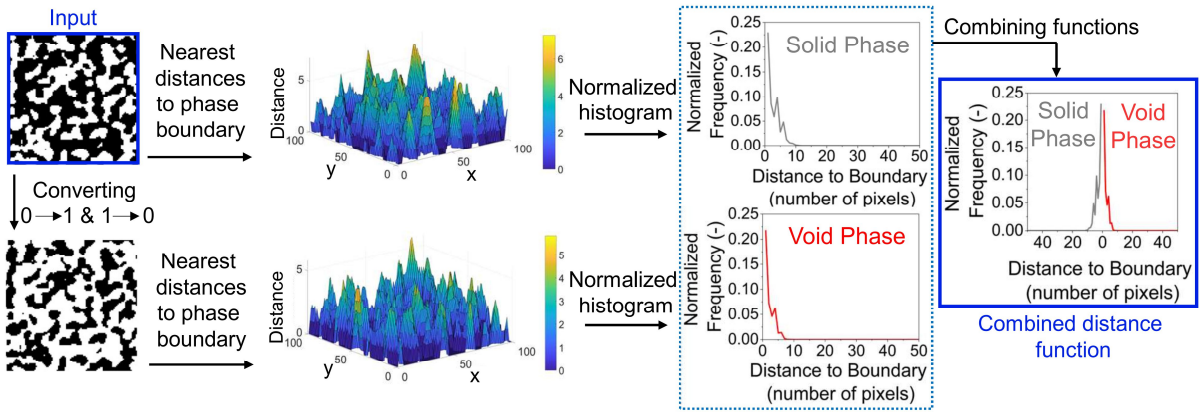

**Fig. 1. Combined distance function for optimization of the membrane structure**. For each 2-D micrograph, the matrix of the nearest distance to the phase boundary (in other words, the distance map) is calculated for both solid and void phases in the form of a histogram. These two histograms are then combined to generate the combined distance function.

## Generating 3-D structures

The next step is to enable rapid and optimizable *in silico* generation of 3-D porous structures. Initially, a 3-D matrix ($A_1$) with the same size as that of the original FIB-SEM structure (1024×899×400 pixels) was generated to represent the membrane. The matrix elements were randomly set to values between 0 and 1. To generate spatial correlation, we filtered $A_1$ with a 3-D Gaussian smoothing kernel (Fig. S4). The standard deviation of the Gaussian filter affects the pore sizes. To account for depth-dependent anisotropy in pore size (i.e., a trend towards increasing pore size with depth as one moves away from the membrane skin layer), a second matrix ($A_2$) was generated with the same dimensions as $A_1$. A second Gaussian filter was used with a larger standard deviation to represent the larger pores occurring towards the bottom of the membrane. To vary pore size with depth, we generated a new matrix ($A_T$) by combining $A_1$ and $A_2$, with relative proportions of the two matrices weighted by the distance between each cell (or pixel) and the top of the 3-D structure (Fig. S5).

Since cells of $A_T$ have values between 0 and 1, thresholding must then be applied on $A_T$ using a cut-off value to binarize the matrix and differentiate between void and solid phases. The cut-off value for this round of thresholding is an important parameter during optimization (Fig. S4). In the next step, morphological closing was used to smooth the sharp corners and make the pores more circular. For this purpose, we used the *imclose* function of MATLAB to perform morphological closing using a structuring element, which defines the pixel in the image being processed and the neighborhood used in the processing of each pixel. The radius of the structuring element controls the level of rounding of the pores; a larger radius in the structural element results in more circularity of the pores. The standard deviations of the Gaussian filters applied on $A_1$ and $A_2$, the thresholding value after forming the weighted average $A_T$, and the radius of the structuring element during morphological closing are the four parameters that determine the final 3-D reconstructed structure.

### Optimization of 3-D structure

To optimize the generated 3-D structure, the algorithm pulls a random cross-sectional slice from the 3-D reconstructed structure (Fig. 2). The combined distance function of this slice can then be determined, and the cumulative absolute difference with the combined distance function of the input micrograph can be calculated. To account for morphological differences in the membrane skin layer (top region near the surface) and the underlying substructure, we divided each image into two sections, the top 20 percent with smaller pores and the bottom 50 percent with larger pores. For each section, a separate combined distance function was generated (Fig. S6). We did not include the middle 30 percent in the model because in this area there was a transition region from small to large pores, which complicated the optimization process.

Compared to other approaches of optimization, Bayesian optimization is a sample-efficient method to find superior solutions for finite and noisy data (37). We used the BayesOpt library of MATLAB (38) to optimize the standard deviations of the Gaussian filter applied on $A_1$ and $A_2$, the thresholding value, and the radius of the structuring element to minimize the cumulative error between the combined distance functions of slices of the 3-D reconstructed structure (for the top and bottom of the image) and the initial 2-D image selected from the original FIB-SEM data set (Fig. 2).

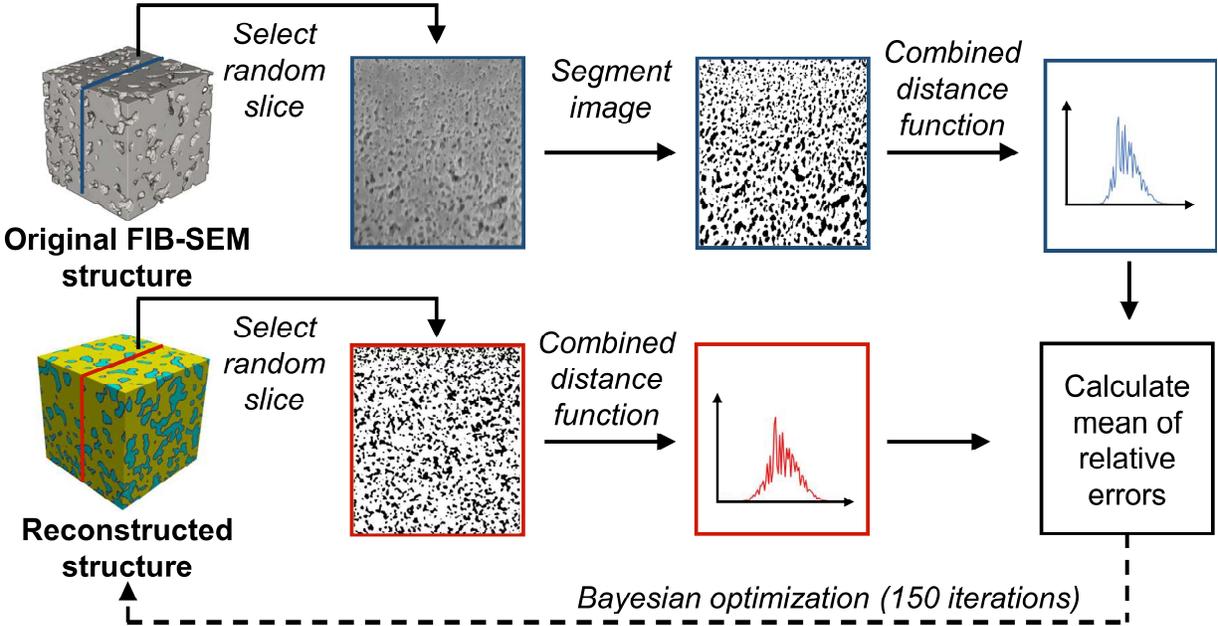

Fig. 2. **Overview of the reconstruction algorithm**. A single slice of the original FIB-SEM is selected randomly, and after image segmentation, the combined distance function is generated. The algorithm generates random 3-D reconstructed structures which are then optimized using Bayesian optimization with the aim of minimizing the error between the combined distance functions of a slice of the reconstructed structure and the one selected from the original FIB-SEM image.

Pore network model

After optimization, a 3-D reconstructed structure is obtained for which a random 2-D slice has a statistically similar morphology compared to the input micrograph according to the combined distance function. Since the input micrograph corresponded to a 3-D FIB-SEM data set, we are also able to extract and compare the properties of the 3-D structures (Fig. 3), allowing us to assess whether optimization using the 2-D image results in statistically similar 3-D structures. The void space in porous materials has previously been studied using pore-network models (39–42). A straightforward pore network model is presented to describe the void/solid phase as a collection of bodies and throats (Fig. S7), with the aim of quantifying the membrane properties. Given that the initial matrix construction was based on random numbers and the slice for comparison was selected randomly, there are some sources of uncertainty in the model. We therefore repeated the model calculations five times.

We first compare the overall properties of the membranes, including the pore radius distribution, throat radius distribution, and coordination number distribution (Fig. 3B-D). The coordination number is the number of connected neighbors for a given pore. For instance, a coordination number of 0 means that the pore is isolated. All three distributions show remarkable similarities between the original and reconstructed structures. Each reconstructed structure is based on a single 2-D micrograph, while the original is based on 400 images, and therefore the agreement seen in Fig. 3B-D is surprising. Table 1 compares the reconstructed and original structures quantitatively. The mean body radius of the original structure is 47.1 nm, while the mean body radius of the reconstructed structures is 40.0 ±1.4 nm. The error bars refer to a 95 percent confidence interval for the expected mean.

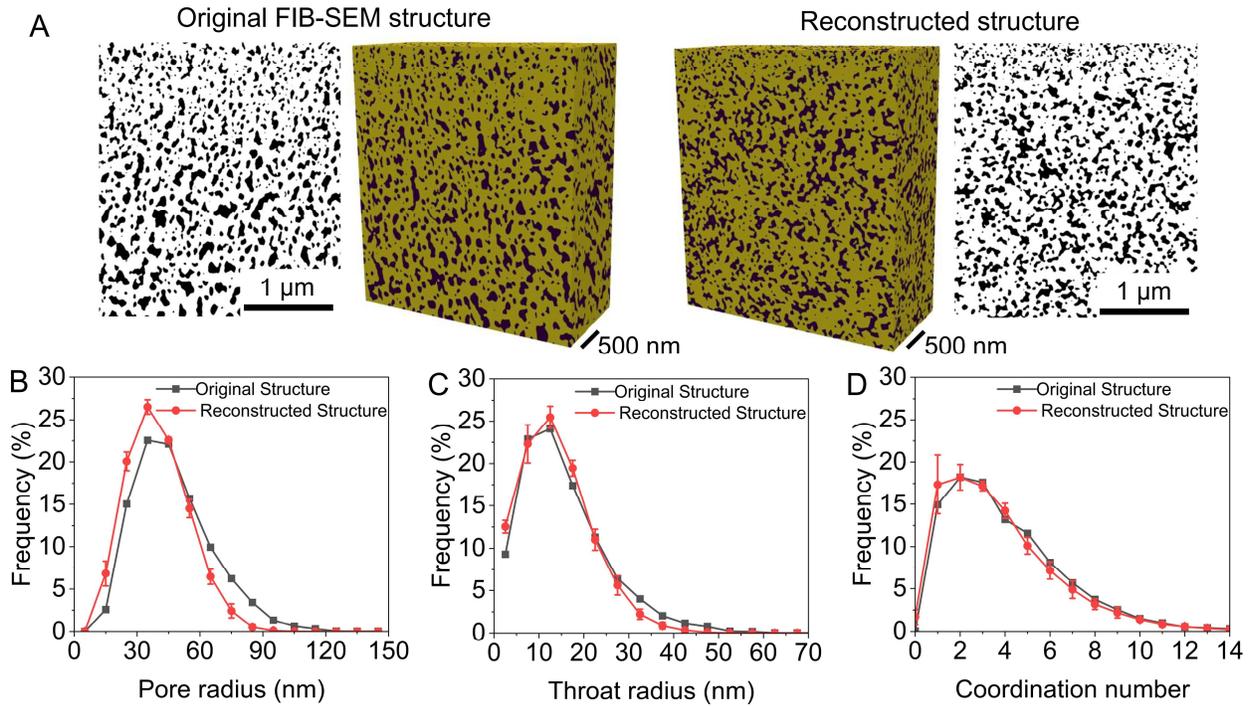

**Fig. 3. Comparison of the original and reconstructed structures**. (A) 2-D and 3-D comparison of the original FIB-SEM and reconstructed structures, and comparison of bulk properties, including (B) body radius distribution, (C) throat radius distribution, and (D) coordination number distribution for the entire structure studied, i.e., the 3 µm depth beginning at the skin layer. 95% confidence interval is calculated for Fig. 3 B-D. Frequency is the ratio of the number of pores in a particular bin to the total number of pores.

The trend of overall porosity (defined as the ratio of the pore to total volume) through the membrane was similar for both the original and reconstructed structures (Fig. S8), showing that the depth-dependent porosity has almost the same range for both structures. Another advantage of having the complete 3-D structure is that the portion of inaccessible pores—i.e., pore area that is completely disconnected from either the reconstruction's inlet and exit surfaces—can be calculated, whereas porometric methods can only assess pores that are connected to both the membrane inlet and exit. Both total and connected porosity are estimated for reconstructed structures with very low errors compared to the original structure. Tortuosity is another membrane parameter affecting mass transfer within the membrane. This parameter is typically described as the ratio between the actual distance traversed between two points and the distance traveled in a straight line connecting the same two points. Therefore, tortuosity can never be less than 1 (43). The developed approach in this study allows us to calculate the tortuosity of the membrane using a single 2-D micrograph with very high accuracy and low standard deviation (Table 1).

Table 1. Comparison between original and reconstructed structures in terms of properties and performance for the entire structure studied.

|  | Mean body radius (nm) | Mean throat radius (nm) | Mean coordination number (-) | Porosity (total) | Porosity (connected) | Tortuosity | Permeability (LMH/psi) |
|---|---|---|---|---|---|---|---|
| Original structure | 47.1 | 15.6 | 4.07 | 0.252 | 0.249 | 1.70 | 13.3 |
| Reconstructed structure #1 | 39.3 | 13.3 | 3.62 | 0.251 | 0.241 | 1.76 | 6.5 |
| Reconstructed structure #2 | 39.8 | 14.1 | 3.75 | 0.261 | 0.256 | 1.73 | 10.7 |
| Reconstructed structure #3 | 37.7 | 12.6 | 3.29 | 0.248 | 0.236 | 1.77 | 5.2 |
| Reconstructed structure #4 | 41.3 | 15.1 | 3.74 | 0.277 | 0.271 | 1.72 | 11.8 |
| Reconstructed structure #5 | 42.1 | 14.3 | 4.62 | 0.276 | 0.274 | 1.72 | 14.3 |
| Average of reconstructions (with 95% confidence interval) | 40.0 ± 1.4 | 13.9 ± 0.8 | 3.80 ± 0.39 | 0.263 ± 0.011 | 0.256 ± 0.013 | 1.74 ± 0.02 | 9.7 ± 3.0 |
| Error (%) | 15 | 10 | 6.6 | 4.3 | 2.8 | 2.3 | 27 |

**Simulations of water and particle transport**

Comprehensive 3-D pore structures can enable numerical approximations of the Navier-Stokes equations via CFD simulations. We used GeoDict to simulate the transport of water passing through the pores of the membrane. Since images were taken from the size-selective skin layer only to a depth of 3 µm, the substructure of the membrane with larger pores were by necessity neglected. The substructure is anticipated to have a low contribution to the resistance to flow because of the fast-expanding pore size trend observed in the membrane structure (44). It has already been shown that the original structure obtained from FIB-SEM is an excellent proxy of the real membrane (20); here our goal is to compare simulated transport in the original FIB-SEM and reconstructed structures. As shown in Table 1, the average estimated water permeability for reconstructed structures is 9.7 LMH/psi, which is 27% lower than the original FIB-SEM structure. This error is greater than those for morphological properties, resulting from water permeability being highly sensitive to the structure of the membrane. For the same reason, the permeabilities of the reconstructed structures vary relatively substantially, with a coefficient of variation of 30% compared to 3.5% for the mean body radius. The variation in simulated water permeability is analogous to experimental permeability measurements using small coupons of the same membrane, for which multiple measurements must be done to determine a meaningful average.

Similar retention properties of gold nanoparticles as those of bacterial and mammalian viruses make them a good proxy for simulating the rejection of virus filters (20). Simulation of the rejection process demonstrates that the reconstructed structures retain gold nanoparticles with 100% efficiency, with a log reduction value (LRV) greater than or equal to 2.7. The original FIB-SEM structure similarly displays 100% rejection, with a LRV greater than or equal to 2.9. The original and reconstructed structures thereby display highly similar rejection properties for 20-nm particles.

## Discussion

Spatial knowledge of the 3-D internal pore structure enables comprehensive structural analysis and exact transport simulations, both of which are impossible when using conventional porometric techniques that only provide pore-size distributions based on a model of the membrane as having an array of parallel pores. Ideally, researchers would be able to do FIB-SEM on each of their samples to obtain these 3-D structures; however, this will never be possible given the complexities and cost of the technique. Our approach is a first step towards the routine characterization of 3-D internal pore structures during membrane development based on readily obtainable 2-D cross-sectional images through the depth of the membrane.

A few key properties enabled the algorithm to successfully achieve our goal of rapidly creating statistically similar 3-D structures using a 2-D input. First, adaptive thresholding enabled robust and precise image segmentation, with the segmented images reflecting void and solid phases even in the presence of shadows (Fig. S1, Table S1). Second, the selected feature (i.e., combined distance function) was not only able to extract significant information embedded in the image—as demonstrated by the similar properties of the original and reconstructed membranes—but also proved to be similar for all slices of the FIB-SEM data set (Fig. S9). This repeatability suggests that selecting a random slice does not introduce meaningful uncertainty to the model, enabling a single micrograph to be used as the model input. The optimization algorithm could successfully find the optimal structure within a few minutes, which is a great achievement compared to FIB-SEM in terms of time.

Future work will aim to improve the algorithm's capture of depth-dependent pore-size variability. While the reconstructed and original structures have similar transport characteristics and overall properties, the depth-dependent pore size distributions differ marginally (Fig. S10). The PSD continues to broaden slightly with depth for the original structure, while for the reconstructed structure, there appears to be two distinct layers.

Our hope is that highly accessible, routine generation of 3-D internal structures will revolutionize the ways that membranes are characterized and modeled. Common transport models used for estimating fluid flow through membrane pores—namely the Hagen–Poiseuille equation, Washburn's equation, and Cantor's equation—assume that membranes comprise discrete cylindrical pores, which dramatically simplifies the highly complex and interconnected nature of true membrane structures such as the one analyzed in this study. CFD simulations of transport through 3-D structures can provide fundamental insights of the impact of structure on transport—for example, enabling tracking of water velocity through the structure (Fig. S11). These simulations could also serve as validation for new mathematical models. Researchers interested in fundamental structure/transport insights in conventional hydraulic flow and in other applications (e.g., membrane distillation (45)) have thus far been limited by the general inability to access relevant 3-D structures.

With further development, a likely advantage of our approach compared to FIB-SEM is the potential to develop a 3-D reconstruction of the pore space through the entire depth of the membrane, which are typically 50–100-μm thick. FIB-SEM analysis can only provide information on the pore space in the 1-5 μm region at the upper surface (inlet region) of the filter, as it is experimentally impractical to mill through much more than a few microns of the membrane. In contrast, it is possible to rapidly generate cross-sectional images through the whole thickness of membranes. These images, in combination with much higher resolution images near the skin, could be used to reconstruct the pore structure throughout the entire thickness of the membrane. Such knowledge would enable fundamental investigation of the impact of the large-pore substructure on transport and mechanical properties, which have largely been neglected to-date, including in our study.

## Materials and Methods

### Membrane FIB-SEM data set

In a previous study (20), the combination of FIB and SEM (FIB-SEM) was used to obtain a 3-D reconstruction of a virus filter (Viresolve® Pro), in which a dual-beam FEI Helios 660 FIB-SEM was utilized for imaging. More details about the operational parameters including voltage, beam current, and tilt angle can be found in the literature (20). After taking images, several steps were done to create 3-D structures of the membrane, namely tilt correction, image alignment, and grayscale adjustment for slice brightness variation. Fig. S12 illustrates a schematic of the 3-D reconstruction using the FIB-SEM technique, carried out in the previous study (20). The data set obtained from this study was used for model development in the current research. This data set was composed of the raw electron micrographs, containing 407 images (or slices) taken at 3-nm slice increments (Data S1). We neglected the last 7 slices owing to some defects found in the last slices.

### Membrane 3-D reconstruction

The 3-D reconstruction algorithm was developed in the MATLAB environment (MathWorks) using the Computer Vision and Statistics and Machine Learning Toolboxes. We repeated the reconstruction process five times, generating five 3-D structures. The confidence interval was calculated for the properties and performance parameters of the membrane. The confidence interval can be calculated by its margin of error. The interval margin of error is dependent on the confidence level that is chosen (46). 95% interval is one of the most common choices that has been used in the literature (47, 48). The margin of error for the 95% confidence interval is calculated as $1.96 \times \frac{SD}{\sqrt{N}}$, where SD is the standard deviation and N is the sample size (46).

### Pore network model

We used the Pore Network Modeling extension of Avizo software (Thermo Fisher Scientific) for spatial analysis. The 3-D structures were initially converted to the .tiff format and then imported to the Avizo environment, in which larger spheres were used to represent pore bodies connected by narrow cylinders, i.e. throats. After thresholding, total porosity was calculated. Axis Connectivity module was then utilized to calculate the connected porosity. For separating pore pathways, we employed the Separate Objects module, which computes watershed lines on a distance map. The Generate Pore Network module was then utilized to calculate PSD, TSD, coordination number distribution (CND), and tortuosity. This module made it possible to retrieve various statistics from a labelled and divided pore space.

**Permeability and rejection**

By employing the FlowDict module of GeoDict (Math2Market, Germany), the water flux through the 3-D structure was assessed by solving the Stokes flow equation at a pressure difference of 210 kPa. Gold nanoparticles have been utilized in several studies to assess the pore size and retention properties of virus filters (44, 49). We used the FilterDict-Media solver to model gold nanoparticle retention under the same pressure as that used to calculate the permeability. At the membrane inlet, feed was supplemented with 20 nm nanoparticles (concentration of $10^8$ particles/mL). Particle-particle and particle-wall collisions, along with Brownian diffusion, were all taken into consideration for nanoparticle transfer through the 3-D structures. When a nanoparticle's diameter is equal to or greater than the pathway accounting for the 3-D structure of the pore space, nanoparticle capture would take place.

Acknowledgments

Funding:

This work was supported by the Acceleration Consortium at the University of Toronto. Funding for this work was also provided by the Natural Sciences and Engineering Research Council of Canada (Alliance Grant ALLRP 570714-2021). KR, AZ, and EG acknowledge support from the Membrane Science, Engineering, and Technology (MAST) Center, which is funded by grant number 1841474 from the U.S. NSF IUCRC program.

Author contributions:

Conceptualization: HC, JH, JW

Model development: HC, AR, JH, JW

Investigation: HC, KR

Visualization: HC, KR, JW

Supervision: AR, EG, AZ, JH, JW

Writing—original draft: HC, AR

Writing—review & editing: EG, AZ, JH, JW

**Competing interests:** Authors declare that they have no competing interests.

**Data and materials availability:** The code is available on GitHub Repository: [Hooman-Chamani/3DReconstruction: Membrane Microstructure Project (github.com)](). The FIB-SEM data set of the virus filter is also present in the paper Supplementary Materials.

# Supplementary Materials for

# Rapid Reconstruction of 3-D Membrane Pore Structure Using a Single 2-D Micrograph

Hooman Chamani, Arash Rabbani, Kaitlyn P. Russell, Andrew L. Zydney, Enrique D. Gomez, Jason Hattrick-Simpers*, Jay R. Werber*

*Corresponding authors. Email: jason.hattrick.simpers@utoronto.ca (J. Hattrick-Simpers) and jay.werber@utoronto.ca (J. Werber)

**This PDF file includes:**

Figs. S1 to S12
Table S1
Data S1

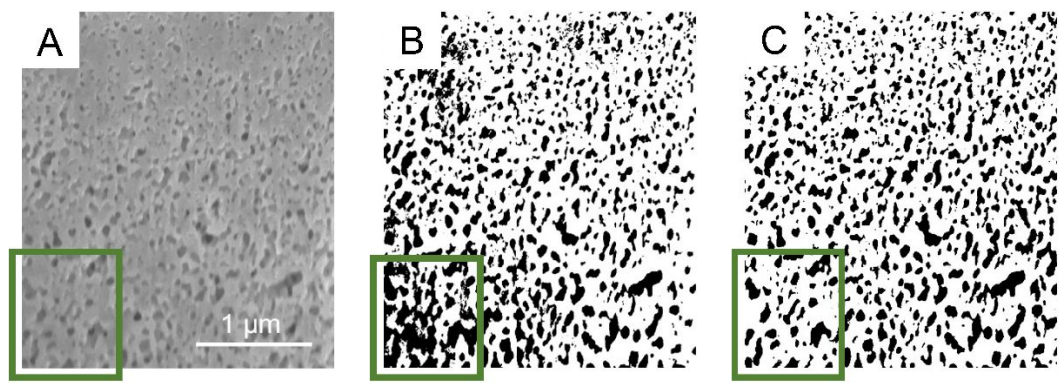

**Fig. S1.**
**Image segmentation.** (A) A middle slice from a FIB-SEM stack that is segmented using (B) conventional thresholding or (C) adaptive thresholding, the latter of which is used in this study. The membrane is oriented with the size-selective skin layer (containing the smallest pores) at the top of the image.

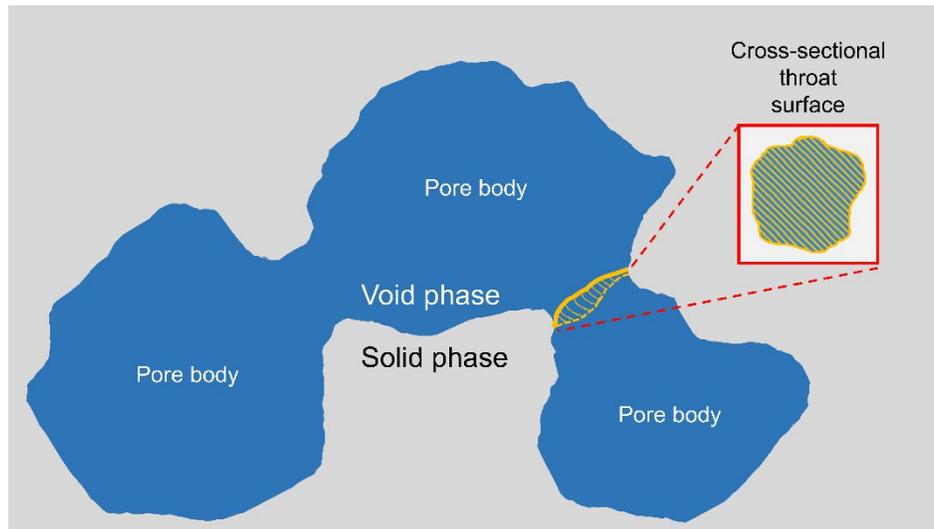

**Fig. S2.**

**Illustration of pore bodies and throats.** This figure displays a schematic of a synthetic porous structure, containing void and solid phases, with few bodies and throats.

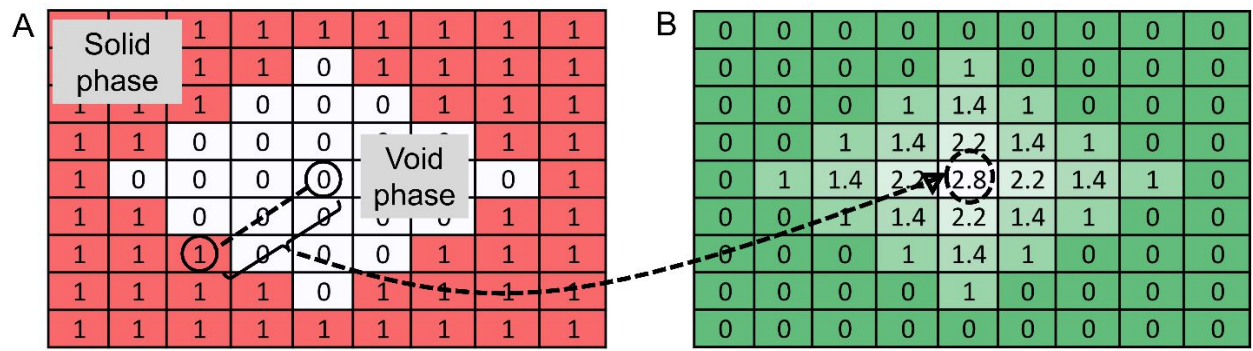

**Fig. S3.**

**Distance map.** (A) A simplified binary image with a single pore and (B) the corresponding distance map (Euclidean distance between each pixel and the nearest non–zero pixel).

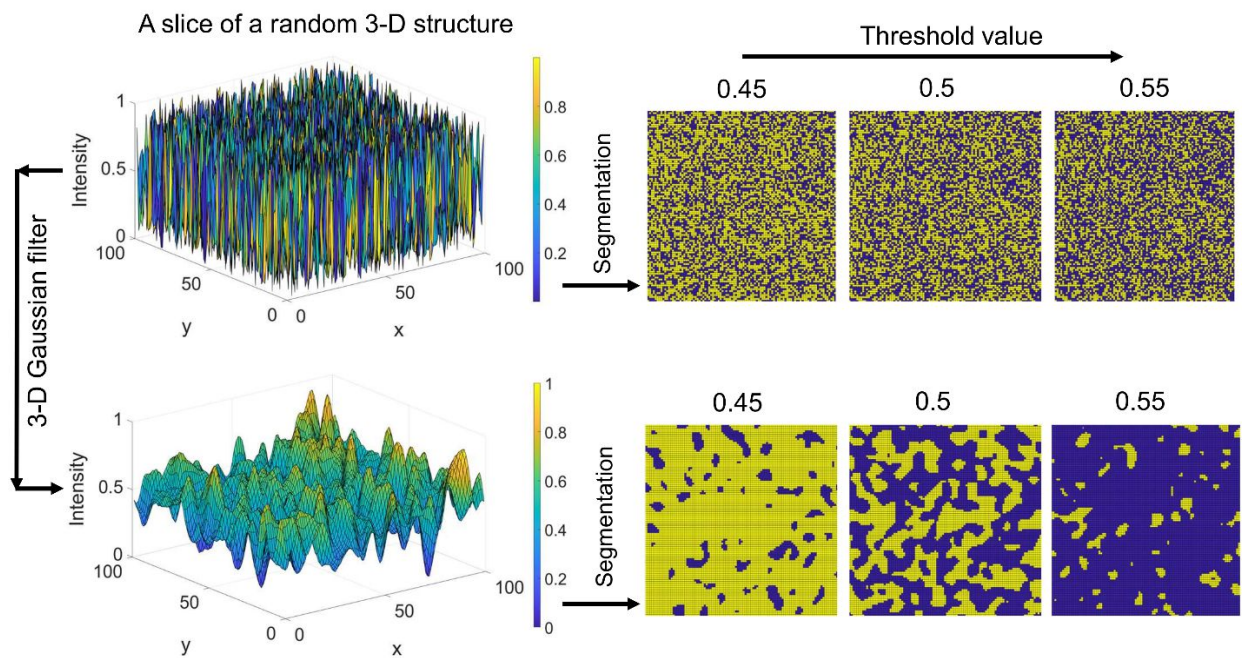

**Fig. S4.**

**3-D Gaussian filter.** The influence of the Gaussian filter on the structure of a synthetic random micrograph with a sigma (standard deviation) of 2.

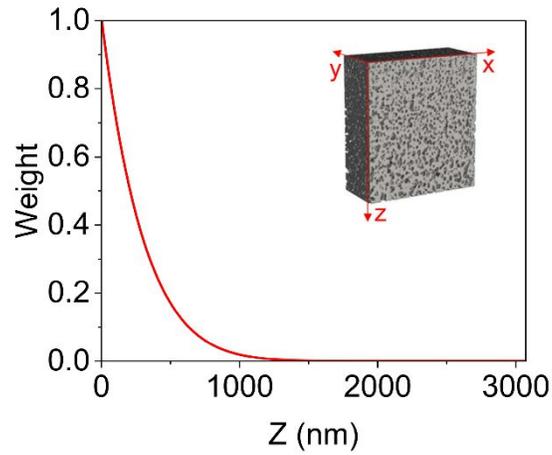

Fig. S5.

**Weight matrix for varying depth-dependent pore sizes.** The weight matrix (W) is a 3-D matrix that changes in the z-direction (z-direction is from the top of the membrane with small pores to the bottom of the membrane with large pores), but is constant in the other two directions (i.e., x and y). This figure was generated for a middle slice in the y-direction. The weight function is given by $(1 - z/h)^{10}$, where $h$ is the total height of the system and $z$ is the distance from the skin layer. The weight function was developed based on visual analysis of pore sizes in the original structure. $A_T$ is calculated using the weighted matrix: $A_T = A_1.*W + A_2.*(1-W)$ (.* is element-wise multiplication). The Gaussian filter applied to $A_1$ and $A_2$ generates a continuous porous structure; the weighted average of the two matrices ($A_T$) similarly has continuous void and solid phases.

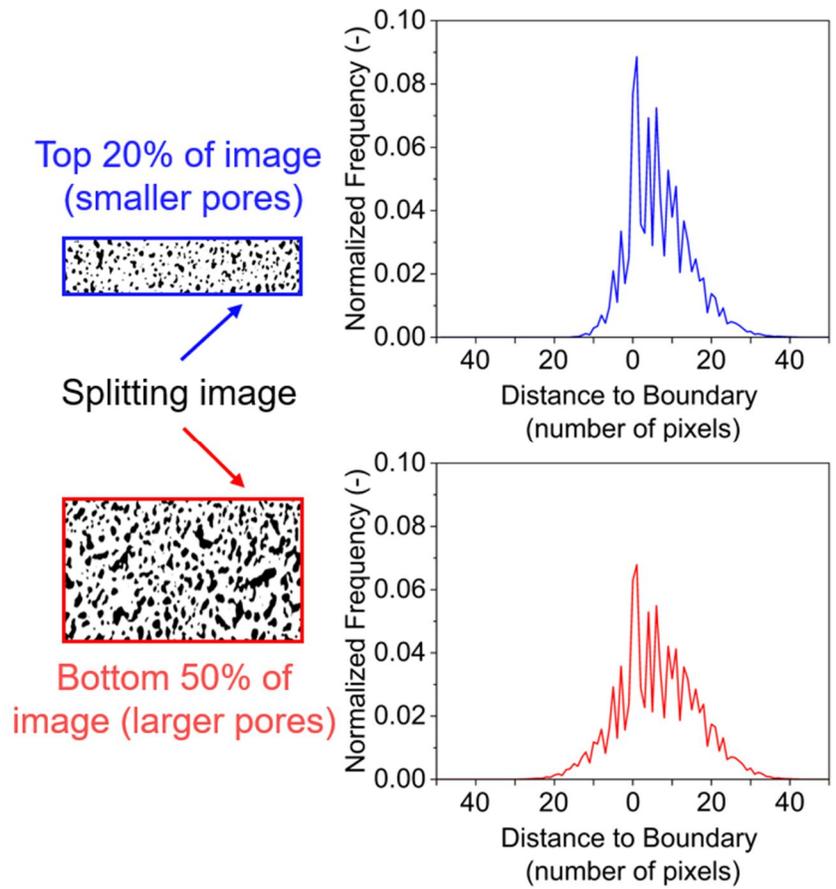

**Fig. S6.**

**Feature extraction from the top and bottom of the image.** Combined distance functions for the top and bottom of a middle slice of the FIB-SEM data set of the Viresolve® Pro filter.

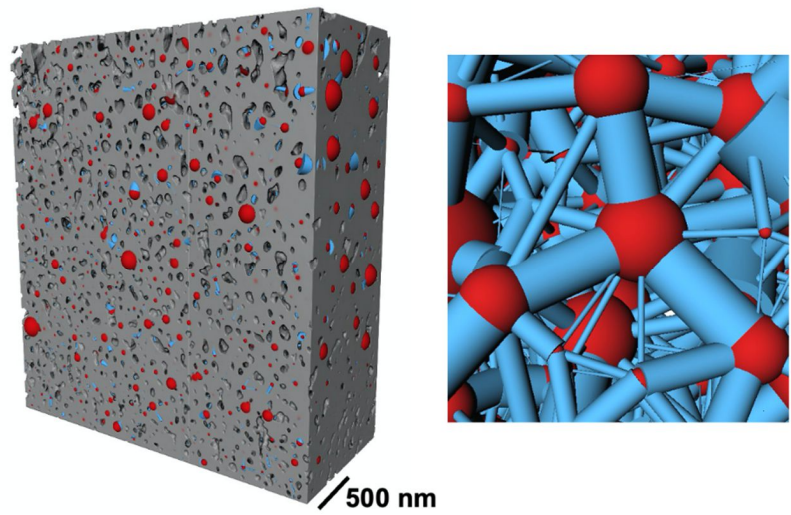

**Fig. S7.**

**Pore network model for a 3-D structure.** This figure is a collection of pores and throats in which red displays bodies and blue illustrates throats.

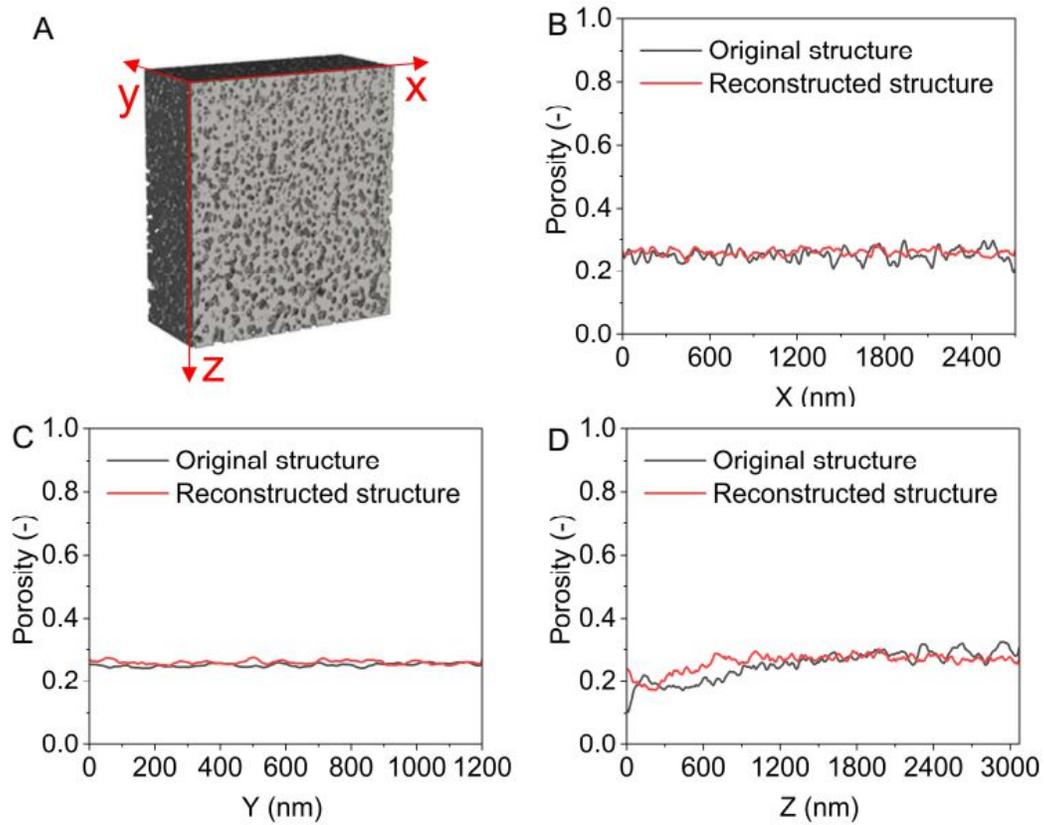

**Fig. S8.**

**Spatial variability of porosity.** Porosity changes for slices taken through the depth of the membrane for both original and reconstructed structures: A) Defining directions, and porosity vs. B) X, C) Y, and D) Z directions.

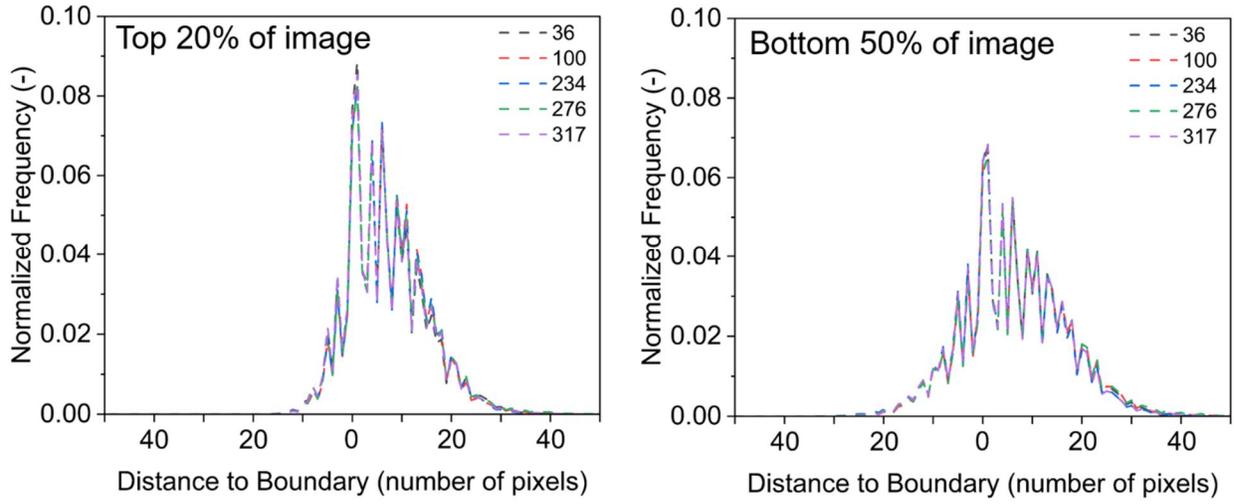

Fig. S9.

**Combined distance functions of the original FIB-SEM structure.** The combined distance function for the top and bottom of some middle slices of the original structure: $36^{th}$, $100^{th}$, $234^{th}$, $276^{th}$ and $317^{th}$ slices.

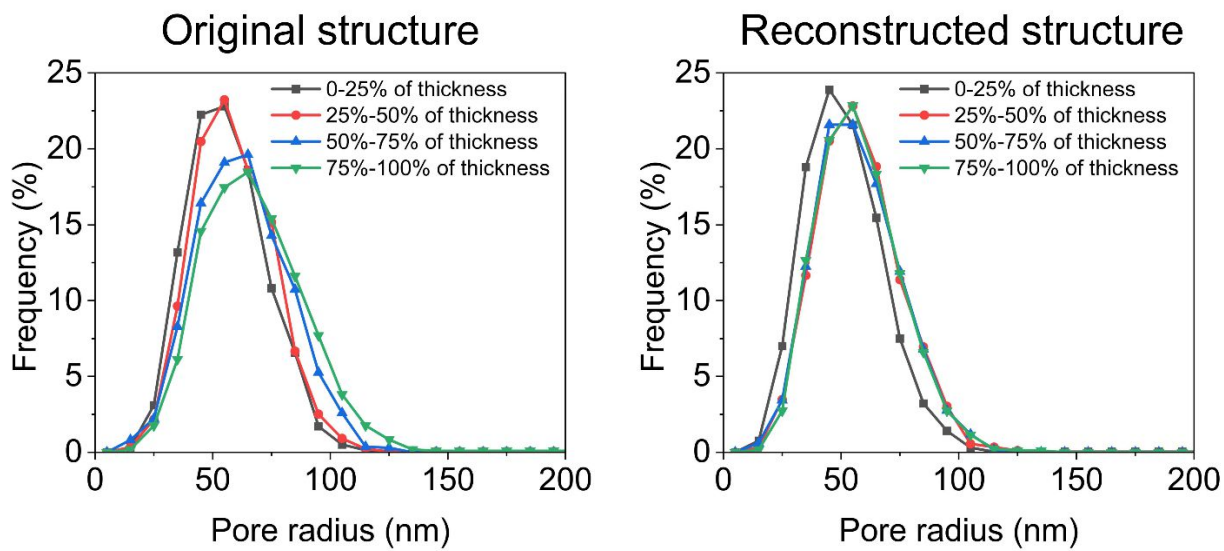

Fig. S10.

**Depth-dependent pore size distribution**. Pore size distribution through the depth of the membrane (in the flow direction, i.e. Z direction, see Fig. S8 for defining directions) for both original FIB-SEM and reconstructed structures, where 0-25% of thickness refers to the uppermost region of the membrane containing the skin layer. Frequency is the ratio of the number of pores in a particular bin to the total number of pores.

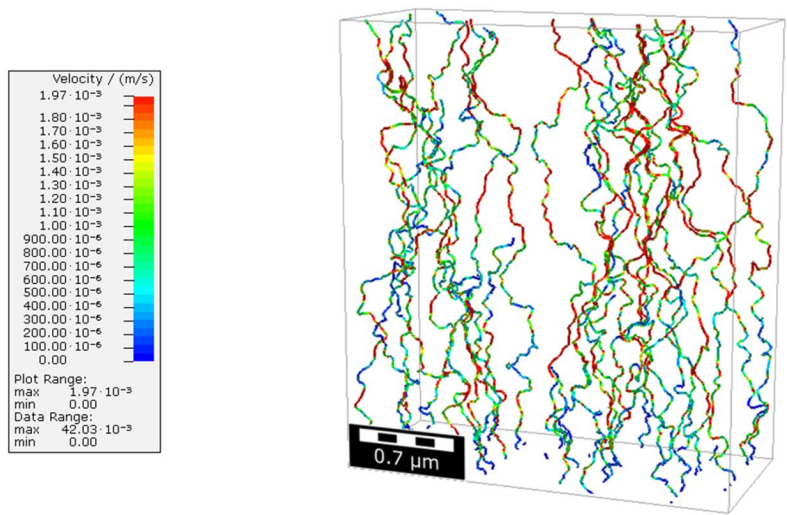

**Fig. S11.**

**Illustration of water flow in the 3-D structure.** This figure is obtained using CFD modeling in GeoDict.

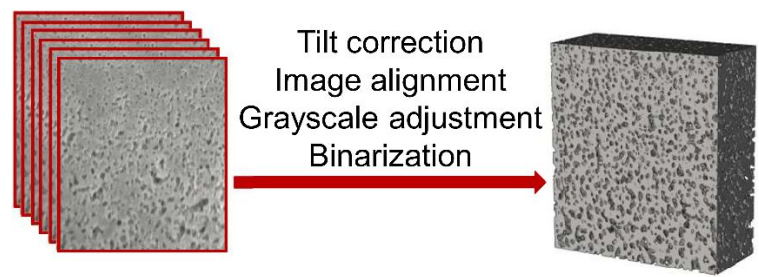

**Fig. S12.**

**3-D reconstruction using FIB-SEM.** The schematic of membrane reconstruction using the FIB-SEM technique.

Table S1. Comparison between original FIB-SEM structures obtained by conventional and adaptive thresholding techniques in terms of properties and performance for the entire structure studied, i.e. 3 µm from the skin layer.

|  | Mean pore radius (nm) | Mean throat radius (nm) | Mean Coordination number (-) | Porosity (total) | Porosity (connected) | Tortuosity | Permeability (LMH/psi) |
|---|---|---|---|---|---|---|---|
| Conventional thresholding (Threshold of 97) | 44.0 | 17.3 | 3.90 | 0.256 | 0.256 | 1.52 | 16.8 |
| Adaptive thresholding | 47.1 | 15.6 | 4.07 | 0.252 | 0.249 | 1.70 | 13.3 |

**Data S1.**

The original FIB-SEM data set of the studied membrane, i.e. Viresolve® Pro filter, is available on the Materials Data Facility.
DOI: 10.18126/FMJE-MYWY
Link: [https://doi.org/10.18126/FMJE-MYWY](https://doi.org/10.18126/FMJE-MYWY)